\documentclass{iopart}
\usepackage{iopams}
\usepackage{amssymb}
\expandafter\let\csname equation*\endcsname\relax
\expandafter\let\csname endequation*\endcsname\relax
\usepackage{amsmath}
\usepackage{bm}
\usepackage{graphicx}
\begin{document}

\title[Ehrenfest theorem in relativistic quantum theory]{Ehrenfest theorem in relativistic quantum theory}
\author{Iwo Bialynicki-Birula}
\address{Center for Theoretical Physics, Polish Academy of Sciences\\
Aleja Lotnik\'ow 32/46, 02-668 Warsaw, Poland\ead{birula@cft.edu.pl}}
\author{Zofia Bialynicka-Birula}
\address{Institute of Physics, Polish Academy of Sciences\\
Aleja Lotnik\'ow 32/46, 02-668 Warsaw, Poland}

\begin{abstract}
Ehrenfest theorem is extended to the relativistic quantum theory of charged particles, moving under the influence of a classical electromagnetic field. In order to extend the original Ehrenfest result to the relativistic domain we bypassed the problems with the relativistic position operator by deriving directly Newton's second law. Our approach is characterized by its universality. The detailed form of the wave equation is not needed. All that is required is the existence of the conserved electric four-current built from the particle wave function. The derivation is based on the conservation laws for the energy and momentum.
\end{abstract}
\submitto{Coherent Control: Photons, Atoms and Molecules\\ The B W Shore Memorial Issue of the Journal of Physics B}

\maketitle
%\ioptwocol

\section{Introduction}

It has been shown by Ehrenfest \cite{pe} that the average momentum of a quantum wave packet obeys the Newtonian equation of classical mechanics. The Ehrenfest theorem illustrates the correspondence between quantum and classical theories and it is mentioned in almost every textbook of quantum mechanics. The Ehrenfest theorem is not really a mathematical theorem because it is just an approximate statement which is valid only ``if the potential changes slowly over the size of a wave packet'' \cite{ap}.

In nonrelativistic quantum mechanics, the derivation of the Ehrenfest equations in \cite{pe} and in all textbooks of quantum mechanics is based on either the Schr\"odinger equation or equivalently on the corresponding equations in the Heisenberg picture. In both cases one relies on the position operator defined as the multiplication by $\bm r$ of the wave function in the position representation. Such a definition is not correct in the relativistic theory because the multiplication by $\bm r$ of the Dirac bispinor describing the electron, will always produce a component of the bispinor describing the positron. We will show that in relativistic quantum theory one may choose a different approach.

The aim of this work is to derive the counterpart of the Ehrenfest theorem in {\em relativistic quantum theory} of charged particles moving under the influence of the electromagnetic field. We shall use the conserved energy-momentum tensor $T^{\mu\nu}(\bm r,t)$ to describe the interaction of a charged relativistic quantum particle with the classical electromagnetic field. In order to associate the motion of the wave packet with the motion of a classical particle, as in the derivation of the nonrelativistic Ehrenfest theorem, we assume that the average charge density $\rho$ is confined to a small region and the velocity $\bm v$ changes slowly over the size of a wave packet. These assumptions obviously restrict the range of physical phenomena that can be correctly described with the use of the Ehrenfest theorem. In most cases (in particular, in the physics of atoms and molecules) this theorem is not applicable.

Unlike the methods used in nonrelativistic quantum mechanics  (cf., for example, \cite{ls}) our derivation, {\em does not require} the knowledge of the specific form of the equation satisfied by the relativistic wave functions: $\phi(\bm r,t),\,\Psi(\bm r,t)$, and $\phi^{\mu}(\bm r,t)$ describing particles with spin 0, spin 1/2, and spin 1. However, for completeness, we write down these equations: the Klein-Gordon-equation, the Dirac equation, and the Proca equation, ($\hbar=1,\,c=1$),
\begin{subequations}\label{weq}
\begin{eqnarray}
D_\mu D^\mu\phi(\bm r,t)+m^2\phi(\bm r,t)=0,\\
\rmi D_\mu\gamma^\mu\Psi(\bm r,t)-m\Psi(\bm r,t)=0,\\
D_\mu\phi^{\mu\nu}(\bm r,t)-m^2\phi^\nu(\bm r,t)=0,
\end{eqnarray}
\end{subequations}
where $D_\mu=\partial_\mu+\rmi eA_\mu$, $\phi^{\mu\nu}=D^\mu\phi^\nu-D^\nu\phi^\mu$ and $A_\mu(\bm r,t)$ is the electromagnetic potential.

The only information that we need for our derivation of the Ehrenfest theorem is that from every solution of the relativistic wave equation one may construct the energy-momentum tensor and the conserved electric four-current density,
\begin{eqnarray}\label{cc}
j^\mu(\bm r,t))=e\{\rho(\bm r,t),\rho(\bm r,t){\bm v}(\bm r,t)\},
\end{eqnarray}
which couples to the electromagnetic field. For the solutions of the Klein-Gordon equation $\phi(\bm r,t)$, the Dirac equation $\Psi(\bm r,t)$, and the Proca equation ${\phi}_\mu(\bm r,t)$, these currents are,
\begin{subequations}\label{cur}
\begin{eqnarray}
j^\mu_{\rm KG}(\bm r,t)&=e{\rm Re}[\phi^*(\bm r,t)\rmi D^\mu\phi(\bm r,t)],\\
j^\mu_{\rm D}(\bm r,t)&=e{\bar\Psi}(\bm r,t)\gamma^\mu\Psi(\bm r,t),\\
j^\mu_{\rm P}(\bm r,t)&=-e{\rm Re}\left[\rmi\phi^*_\nu(\bm r,t)\,\phi^{\mu\nu}(\bm r,t)\right].
\end{eqnarray}
\end{subequations}

It is worth noticing, that the charge density $\rho(\bm r,t)$ is positive definite only for the solutions of the Dirac equation. In the case of the Klein-Gordon and Proca equations, the charge density may change its sign in some regions.

To avoid any misinterpretation of our results, we summarize the basic tenets of our approach which closely follow {\em mutatis mutandis} the original ideas of Ehrenfest.\\
1. We consider the solutions of relativistic wave equations in the presence of a given electromagnetic field.\\
2. Out of these solutions we select only those that for some period of time are confined to a region which is small as compared to the characteristic scale of the field change.\\
3. The form of the equations of motion of the center of the wave packet does not depend on whether the wave function describes the particle, the antiparticle, or the combination of both. However, the interpretation and the applicability to a realistic physical situation requires the decision whether we want to describe the trajectory of a particle or an antiparticle. This decision must be made before we fix the sign of $e$ in the definition of the charge density in (\ref{cc}). This sign determines whether the wave packet describes predominantly the particle or the antiparticle. When both components are present, the choice of sign is ambiguous and the physical interpretation of the trajectory as describing the motion of either a particles or an antiparticle is impossible.\\
4. The velocity of the wave packets is defined in terms of the current (\ref{cc}).\\
5. The equations of motion for the center of the wave packet are derived from the conservation laws of the energy-momentum tensor.

\section{The energy-momentum tensor}

Let us consider a system composed of a relativistic particle, the classical electromagnetic field and some unspecified external sources of this field. The total energy-momentum tensor has three parts: $T^{\mu\nu}$ for the particle part, $T^{\mu\nu}_F$ for the field part, and $T^{\mu\nu}_E$ for the external sources of the field. For a closed isolated system, the total energy-momentum tensor must be conserved; it satisfies the continuity equation,
\begin{eqnarray}\label{em}
 \partial_\mu \left[T^{\mu\nu}(\bm r,t)+T^{\mu\nu}_{\rm F}(\bm r,t)+T^{\mu\nu}_{\rm E}(\bm r,t)\right]=0.
\end{eqnarray}
The energy-momentum tensor $T^{\mu\nu}(\bm r,t)$ for relativistic quantum particles is built from their wave functions. The expressions for the energy-momentum tensors were obtained for spin 0 particles by Schr\"odinger \cite{es}, for spin 1/2 particles by Tetrode \cite{tet}, and for spin 1 particles by Proca \cite{proca}. We list below these expressions,
\begin{subequations}\label{3t}
\begin{eqnarray}
T^{\mu\nu}_{\rm KG}&={\rm Re} \left[(D^\mu\phi)^*D^\nu\phi
\right]
-g^{\mu\nu}L_{\rm KG}\\
T^{\mu\nu}_{\rm D}&=\frac{1}{2}{\rm Re} \left[{\bar\Psi}\gamma^\mu\rmi \partial^\nu\Psi+{\bar\Psi}\gamma^\nu \rmi\partial^\mu\Psi\right]-g^{\mu\nu}L_{\rm D},\\
T^{\mu\nu}_{\rm P}&={\rm Re} \left[-\phi^{*\mu}_{\;\;\;\lambda}\phi^{\nu\lambda}
+m^2\phi^{*\mu}\phi^{\nu}\right]-g^{\mu\nu}L_{\rm P},
\end{eqnarray}
\end{subequations}
where
\begin{subequations}\label{3l}
\begin{eqnarray}
L_{\rm KG}&=\frac{1}{2}\left(D_\lambda\phi)^*
D^\lambda\phi-m^2\phi^*\phi\right),\\
L_{\rm D}&={\rm Re} \left[{\bar\Psi}\gamma^\mu \rmi\partial_\mu\Psi\right]-m{\bar\Psi}\Psi,\\
L_{\rm P}&=-\frac{1}{4}\phi^{*\mu\nu}
\phi_{\mu\nu}+\frac{1}{2}m^2\phi^{*\mu}\phi_{\mu}.
\end{eqnarray}
\end{subequations}
For brevity we omitted the space-time arguments of all wave functions.

The coupling of the particle and the sources to the field causes, of course, the exchange of energy and momentum between the electromagnetic field and the remaining parts of the system.

Charged sources couple to the electromagnetic field through their charge density and current density, appearing in Maxwell equations,
\begin{eqnarray}\label{cons0}
\partial_t{\bm D}-\nabla\!\times\!{\bm H}=-\rho{\bm v}-\rho_{\rm E}{\bm v}_{\rm E},\\
\nabla\!\cdot\!{\bm D}=\rho+\rho_{\rm E},\\
\partial_t{\bm B}+\nabla\!\times\!{\bm E}=0,\\
\;\nabla\!\cdot\!{\bm B}=0,
\end{eqnarray}
where ${\bm D}=\epsilon {\bm E}$ and ${\bm B}=\mu {\bm H}$.

The components of the energy-momentum tensor of the electromagnetic field are \cite{jdj,qed}:\\
the energy density,
\begin{eqnarray}\label{en}
T^{00}_{\rm F}=\textstyle{\frac{1}{2}}\left(\bm D\!\cdot\!\bm E+\bm B\!\cdot\!\bm H\right),
\end{eqnarray}
the momentum density,
\begin{eqnarray}\label{mom}
\frac{1}{c}T^{0i}_{\rm F}=\left(\bm D\!\times\!\bm B\right)^i.
\end{eqnarray}
and the Maxwell stress-tensor,
\begin{eqnarray}\label{mst}
T^{ij}_{\rm F}=-D^iE^j-B^iH^j+\textstyle{\frac{1}{2}}
\delta^{ij}\left(\bm D\!\cdot\!\bm E+\bm B\!\cdot\!\bm H\right).
\end{eqnarray}
The time derivatives of the energy density and momentum density can be calculated with the use of the Maxwell equations,
\begin{eqnarray}\label{tder}
\partial_t T^{00}_{\rm F}=-\rho{\bm v}\!\cdot\!{\bm E}-\rho_{\rm E}{\bm v}_{\rm E}\!\cdot\!{\bm E}-\nabla\!\cdot\!({\bm E}\!\times\!{\bm H}),\\
\partial_t T^{0i}_{\rm F}=-\rho\left({\bm E}+{\bm v}\!\times\!{\bm B}\right)^i-\rho_{\rm E}\left({\bm E}+{\bm v}_{\rm E}\!\times\!{\bm B}\right)^i-\partial_kT^{ki}_{\rm F}.
\end{eqnarray}
Integrating both sides of these equations over the sufficiently large space region $V$ with the boundary surface $S$ that encloses the wave packet and the field sources, we obtain,
\begin{eqnarray}
\fl\frac{d}{dt}\int_V\!dV T^{00}_{\rm F}=-\int_V\!dV\,\rho\,{\bm v}\!\cdot\!{\bm E}-\int_V\!dV\,\rho_{\rm E}{\bm v}_{\rm E}\!\cdot\!{\bm E}\nonumber\\
-\oint_S dS\,{\bm n}\!\cdot\!({\bm E}\!\times\!{\bm H}),\label{cons1}\\
\fl\frac{d}{dt}\int_V\!dV T^{0i}_{\rm F}=-\int_V\!dV\,\rho\,({\bm E}+{\bm v}\!\times\!{\bm B})^i\nonumber\\
-\int_V\!dV\,\rho_{\rm E}({\bm E}+{\bm v}_{\rm E}\!\times\!{\bm B})^i-\oint_S dS\,n_kT^{ki}_{\rm F}.\label{cons2}
\end{eqnarray}
Each term in these equations has a clear physical interpretation. The first term in Eq.(\ref{cons1}) describes the energy exchanged between the electromagnetic field and the particle part of the system. The second term in this equation describes the energy exchanged between the electromagnetic field and the field sources. Finally, the last term, obtained with the use of the Gauss-Ostrogradsky theorem, describes the energy carried by the radiation of electromagnetic waves through a closed surface $S$. The interpretation of Eq.(\ref{cons2}) is analogous to Eq.(\ref{cons1}), with the energy replaced by the momentum. We are only interested in the terms that describe the exchange of the energy and momentum between the electromagnetic field and the particle which are described by the first terms in both equations. The remaining terms play no role in the derivation of the relativistic Ehrenfest equations.

\section{Relativistic Ehrenfest theorem}

The momentum of the classical particle can be calculated from the components $T^{0i}(\bm r,t)$ of the energy-momentum tensor for the particle. However, we do not need the explicit form of this tensor. It is sufficient to know that such formulas exist and could be used to define the particle energy $\mathcal E$ and momentum $\mathcal P$,
\begin{eqnarray}\label{ep}
{\mathcal E}(t)=\int_V dV T^{00}(\rm r,t),\\ {\mathcal P}^i(t)=\int_V dV T^{0i}(\rm r,t).
\end{eqnarray}
In turn, we determine the amount of energy and momentum per unit time which are exchanged between the electromagnetic field and the particle from Eqs.~(\ref{cons1}) and (\ref{cons2}). Using this information, we obtain,
\begin{eqnarray}
\frac{d}{dt}{\mathcal E}(t)&\!=\!\int_V\!dV\,\rho\,({\bm r},t){\bm v}({\bm r},t)\!\cdot\!{\bm E}({\bm r},t),\label{ehr1}\\
\frac{d}{dt}{\bm{\mathcal P}}(t)&\!=\!\int_V\!dV\,\rho\,({\bm r},t)\left[{\bm E}({\bm r},t)+{\bm v}({\bm r},t)\!\times\!{\bm B}({\bm r},t)\right].\label{ehr2}
\end{eqnarray}

Of course, the same result for the energy and the momentum exchanged between the field and the particle can be obtained by evaluating directly the corresponding time derivatives of the integrated components of the energy-momentum tensors of the particles given by Eqs.(\ref{3t}). This calculation is relatively simple for the spin 0 particle but quite involved for spin 1/2 and 1. The advantage of our method is that we obtain at once the results for all kinds of particles without doing any cumbersome calculations. This is done just by observing what is happening with the energy of the electromagnetic field, as described by Eqs.(\ref{cons1}) and (\ref{cons2}).

We shall now invoke the key assumptions that in a certain period of time the size of the wave packet is very small as compared to the parameters that characterize the space dependence of the electromagnetic field and of the wave packet velocity. This means that in Eqs.~(\ref{ehr1}) and (\ref{ehr2}) we may replace the values of the electric and magnetic field vectors at the point $\bm r$ by their values at the center of the charge density denoted by $\bm{\xi}(t)$. The remaining integral of $\rho$ over $\bm r$ produces the total charge,
\begin{eqnarray}\label{delta}
\int_V\!dV\,\rho\,=e.
\end{eqnarray}
This approximation leads to the equations of motion that have the same form as those for a classical charged particle,
\begin{eqnarray}
\frac{d}{dt}{\mathcal E}(t)&=e{\bm v}(t)\!\cdot\!{\bm E}({\bm\xi}(t),t),\label{ehr3}\\
\frac{d}{dt}{\bm{\mathcal P}}(t)&=e({\bm E}({\bm\xi}(t),t)+{\bm v}(t)\!\times\!{\bm B}({\bm\xi}(t),t)),\label{ehr4}
\end{eqnarray}
where ${\bm v}(t)={\bm v}({\bm\xi}(t),t)$ is the velocity at the position ${\bm\xi}(t)$.

One may also calculate corrections to the relativistic Ehrenfest theorem, as has been done by Shankar in the nonrelativistic case \cite{rs}, by expanding the electromagnetic field into the Taylor series around the point ${\bm\xi}(t)$. The lowest order correction $\bm F^{(1)}_L$ to the Lorentz force appearing in (\ref{ehr4}), obtained by keeping only the term with the first derivatives in the Taylor series, is
\begin{eqnarray}\label{corr}
\bm F^{(1)}_L(t)=e d^i(t)\frac{\partial}{\partial x^i}({\bm E}({\bm r},t)+{\bm v}(t)\!\times\!{\bm B}({\bm r},t))\Big |_{{\bm r}\to{\bm\xi}(t)},
\end{eqnarray}
where ${\bm d}(t)$ is the first moment of the charge distribution with respect to its center,
\begin{eqnarray}\label{1mom}
{\bm d}(t)=\int dV\,{\bm r}\rho({\bm\xi}(t)+{\bm r},t).
\end{eqnarray}
This correction is clearly of the order of $l/L$, where $l$ measures the size of wave-packet and $L$ characterizes the scale of changes of the electromagnetic field. For a spherically symmetric wave the first order correction vanishes and only the second correction proportional to the quadrupole moment of the charge distribution will contribute.

This ends our derivation of the relativistic version of the Ehrenfest theorem. We have shown that the center of the wave packet associated with the relativistic particle (say an electron) obeys Newton's second law expressed in terms of momentum. The force has the form of the Lorentz force. The same expression appears in the nonrelativistic version of the Ehrenfest theorem \cite{ls}. In that case, however, on the l.h.s. we have the acceleration multiplied by the mass which in the relativistic case is not the same as the time derivative of the momentum. The formulation of Newton's second law in terms of momentum is universal. It has the same form in both cases: in the nonrelativistic and in the relativistic theory with the same expression for the Lorentz force. On this occasion we should admire Newton's foresight who formulated his second law in a way that has not required modifications in the relativistic theory.

We believe that our results might also contribute to the discussion started by Einstein and Laub \cite{el} and still continued \cite{mm0,mm} on the
incompatibility of the Lorentz force with special relativity. In particular, we have shown that in the case of relativistic quantum particles there is no need to take into account some ``hidden momentum'' \cite{ws}. However, our results apply only to the motion in free space and not to the motion in the ponderable medium, as in the case treated in \cite{mm0}.

\end{document}